\documentclass[fleqn,twoside]{article}
\usepackage[headings]{espcrc2_my}

\readRCS
$Id: espcrc2.tex,v 1.2 2004/02/24 11:22:11 spepping Exp $
\usepackage{graphicx}
\usepackage{pslatex}
\usepackage[latin1]{inputenc}
\usepackage[T1]{fontenc}

\def\bq{\begin{eqnarray}}
\def\eq{\end{eqnarray}}

\def\eps{\varepsilon}

\title{Infrared finite cross sections at NNLO}

\author{  S.~Weinzierl
 \address{Max-Planck-Institut f\"ur Physik 
    (Werner-Heisenberg-Institut), F\"ohringer
    Ring 6, D-80805 M\"unchen, Germany}  
 \thanks{Heisenberg fellow of the
    Deutsche Forschungsgemeinschaft}
}
       
\runtitle{Infrared finite cross sections}
\runauthor{S. Weinzierl}

\begin{document}

\begin{abstract}
I discuss methods for the cancellation of infrared divergences at NNLO.
\end{abstract}

\maketitle

\section{Introduction}
\label{sec:intro}

In recent years there has been significant progress in the calculation of two-loop amplitudes
\cite{Bern:2000ie,Bern:2000dn,Anastasiou:2000kg,Anastasiou:2000ue,Anastasiou:2000mv,Anastasiou:2001sv,Glover:2001af,Bern:2001dg,Bern:2001df,Bern:2002tk,Garland:2001tf,Garland:2002ak,Moch:2002hm}.
These amplitudes are needed for fully differential 
next-to-next-to-leading order (NNLO) calculations to improve
the accuracy of theoretical predictions relevant to high-energy collider experiments.
At the next-to-next-to-leading order level the ingredients for the third order term in the perrturbative
expansion for quantities depending on $n$ resolved ``hard'' partons
are the already mentionend $n$-parton two-loop amplitudes, the $(n+1)$-parton one-loop amplitudes
and the $(n+2)$ Born amplitudes.
Taken separately, each one of these contributions is infrared divergent. Only the sum of all contributions
is infrared finite.
Here I review the state of the art for the cancellation of infrared divergences between these different
contributions at NNLO.

Infrared divergences occur already at next-to-leading order.
At NLO real and virtual corrections contribute.
The virtual corrections contain the loop integrals and can have,
in addition to ultraviolet divergences, infrared divergences.
If loop amplitudes are calculated in dimensional regularisation,
the IR divergences manifest themselves as
explicit poles in the 
dimensional regularisation parameter $\varepsilon=2-D/2$.
These poles cancel with similar poles arising from
amplitudes with additional partons but less internal loops, when integrated over phase space regions where
two (or more) partons become ``close'' to each other.
In general, the Kinoshita-Lee-Nauenberg theorem
\cite{Kinoshita:1962ur,Lee:1964is}
guarantees that any infrared-safe observable, when summed over all 
states degenerate according to some resolution criteria, will be finite.
However, the cancellation occurs only after the integration over the unresolved phase space
has been performed and prevents thus a naive Monte Carlo approach for a fully exclusive
calculation.
It is therefore necessary to cancel first analytically all infrared divergences and to use
Monte Carlo methods only after this step has been performed.
At NLO, general methods to circumvent this problem are known.
This is possible due to the universality of the singular behaviour
of the amplitudes in soft and collinear limits.
Examples are the phase-space slicing method
\cite{Giele:1992vf,Giele:1993dj,Keller:1998tf}
and the subtraction method
\cite{Frixione:1996ms,Catani:1997vz,Dittmaier:1999mb,Phaf:2001gc,Catani:2002hc}.

I briefly review the subtraction method here.
Within the subtraction method 
one subtracts a suitable approximation term $d\sigma^A$ 
from the real corrections $d\sigma^R$.
This approximation term must have the same singularity structure as the real corrections.
If in addition the approximation term is simple enough, such that it can be integrated analytically
over a one-parton subspace, then the result can be added back to the virtual corrections $d\sigma^V$.
\begin{eqnarray}
\sigma^{NLO} & = & 
\int\limits_{n+1} \left( d\sigma^R - d\sigma^A \right) 
+ \int\limits_n \left( d\sigma^V + \int\limits_1 d\sigma^A \right). 
\nonumber
\end{eqnarray}
Since by definition $d\sigma^A$ has the same singular behaviour as $d\sigma^R$, $d\sigma^A$
acts as a local counter-term and the combination $(d\sigma^R-d\sigma^A)$ is integrable
and can be evaluated numerically.
Secondly, the analytic integration of $d\sigma^A$ over the one-parton subspace will yield
the explicit poles in $\varepsilon$ needed to cancel the corresponding poles in $d\sigma^V$.
The simplest example are 
the NLO corrections to $\gamma^\ast \rightarrow 2 \; \mbox{jets}.$
The real corrections are given by the matrix element for 
$\gamma^\ast \rightarrow q(p_1) g(p_2) \bar{q}(p_3)$ and read, 
up to colour and coupling factors
\begin{eqnarray}
\lefteqn{
 \left| {\cal A}_3 \right|^2 = 8 ( 1 - \varepsilon)  } 
 & & \nonumber \\
 & & 
 \left[
         \frac{2}{x_1 x_2} 
         - \frac{2}{x_1}
         - \frac{2}{x_2} 
         + (1-\varepsilon) \frac{x_2}{x_1}
         + (1-\varepsilon) \frac{x_1}{x_2} 
         - 2 \varepsilon 
        \right],
 \nonumber
\end{eqnarray}
where $x_1=s_{12}/s_{123}$ and $x_2=s_{23}/s_{123}$.
Singularities occur at the boundaries of the integration region at $x_1=0$ and $x_2=0$.
The approximation term can be taken as a sum of two (dipole) subtraction
terms:
\begin{eqnarray}
\lefteqn{
d\sigma^A
 = 
\left| {\cal A}_2(p_1',p_3') \right|^2 \frac{1}{s_{123}}
       \left[
         \frac{2}{x_1 (x_1 + x_2)} - \frac{2}{x_1}
\right.
} & & \nonumber \\
 & & 
\left.
         + (1-\varepsilon) \frac{x_2}{x_1}
        \right]
 +
\left| {\cal A}_2(p_1'',p_3'') \right|^2 \frac{1}{s_{123}}
 \left[
         \frac{2}{x_2 (x_1 + x_2)} 
\right.
 \nonumber \\
 & &
\left.
         - \frac{2}{x_2}
         + (1-\varepsilon) \frac{x_1}{x_2}
        \right].
 \nonumber 
\end{eqnarray}
The momenta $p_1'$, $p_3'$, $p_1''$ and $p_3''$ are linear combinations of the original momenta $p_1$, $p_2$ and $p_3$.
The first term is an approximation for $x_1 \rightarrow 0$, whereas the second term is an approximation
for $x_2 \rightarrow 0$.
Note that the soft singularity is shared between the two dipole terms 
and that in general the Born amplitudes ${\cal A}_2$ are evaluated with different momenta.
The subtraction terms can be derived by working in the axial gauge. In this gauge only diagrams where the emission
occurs from external lines are relevant for the subtraction terms.
Alternatively, they can be obtained from off-shell currents and
antenna factorisation 
\cite{Kosower:1998zr,Kosower:2002su,Kosower:2003cz,Kosower:2003bh}.

Once suitable subtraction terms are found, they have to be integrated over the unresolved phase space.
Here, one faces integrals with overlapping divergences, as one can already see from our simple example:
\begin{eqnarray}
\lefteqn{
     \int d^3 x \delta\left( 1 - \sum\limits_{i=1}^3 x_i \right) 
          x_1^{-\varepsilon} x_2^{-\varepsilon} x_3^{-\varepsilon}
} & & \nonumber \\
 & &
       \left[
         \frac{2}{x_1 (x_1 + x_2)} - \frac{2}{x_1}
         + (1-\varepsilon) \frac{x_2}{x_1}
        \right]
\label{nlo}     
\end{eqnarray}
The term $1/(x_1+x_2)$ is an overlapping singularity. Sector decomposition 
\cite{Hepp:1966eg,Roth:1996pd,Binoth:2000ps,Binoth:2004jv}
is a convenient tool to disentangle overlapping singularities.
Other techniques make use of the optical theorem to convert phase space integrals into loop integrals
\cite{Anastasiou:2003gr}. 
First applications of these methods to processes like $e^+ e^- \rightarrow \mbox{2 jets}$
or $p p \rightarrow W$ have become available
\cite{Gehrmann-DeRidder:2003bm,Gehrmann-DeRidder:2004tv,Anastasiou:2003ds,Anastasiou:2004qd,Kilgore:2004ty}.

\section{The subtraction method at NNLO}
\label{sect:doubleunres}

The following terms contribute at NNLO:
\begin{eqnarray}
\lefteqn{
d\sigma_{n+2}^{(0)} =  
 \left( \left. {\cal A}_{n+2}^{(0)} \right.^\ast {\cal A}_{n+2}^{(0)} \right) 
d\phi_{n+2}, } \nonumber \\
\lefteqn{
d\sigma_{n+1}^{(1)} =  
 \left( 
 \left. {\cal A}_{n+1}^{(0)} \right.^\ast {\cal A}_{n+1}^{(1)} 
 + \left. {\cal A}_{n+1}^{(1)} \right.^\ast {\cal A}_{n+1}^{(0)} \right)  
d\phi_{n+1}, } \nonumber \\
\lefteqn{d\sigma_n^{(2)} = } & & 
 \nonumber \\
 & & 
 \left( 
 \left. {\cal A}_n^{(0)} \right.^\ast {\cal A}_n^{(2)} 
 + \left. {\cal A}_n^{(2)} \right.^\ast {\cal A}_n^{(0)}  
 + \left. {\cal A}_n^{(1)} \right.^\ast {\cal A}_n^{(1)} \right) d\phi_n, \nonumber 
\end{eqnarray}
where ${\cal A}_n^{(l)}$ denotes an amplitude with $n$ external partons and $l$ loops.
$d\phi_n$ is the phase space measure for $n$ partons.
We would like to construct a numerical program for an arbitrary infrared safe observable ${\cal O}$.
Infrared safety implies that whenever a 
$n+l$ parton configuration $p_1$,...,$p_{n+l}$ becomes kinematically degenerate 
with a $n$ parton configuration $p_1'$,...,$p_{n}'$
we must have
\begin{eqnarray}
{\cal O}_{n+l}(p_1,...,p_{n+l}) & \rightarrow & {\cal O}_n(p_1',...,p_n').
\end{eqnarray}
To render the individual contributions finite, one adds and subtracts suitable
pieces
\cite{Weinzierl:2003fx,Weinzierl:2003ra}:
\begin{eqnarray}
\lefteqn{
\langle {\cal O} \rangle_n^{NNLO} = } & &
 \nonumber \\
 & &
 \int 
               {\cal O}_{n+2} \; d\sigma_{n+2}^{(0)} 
             - {\cal O}_{n+1} \circ d\alpha^{(0,1)}_{n+1}
             - {\cal O}_{n} \circ d\alpha^{(0,2)}_{n} 
 \nonumber \\
& &
 + \int 
                 {\cal O}_{n+1} \; d\sigma_{n+1}^{(1)} 
               + {\cal O}_{n+1} \circ d\alpha^{(0,1)}_{n+1}
               - {\cal O}_{n} \circ d\alpha^{(1,1)}_{n}
 \nonumber \\
& & 
 + \int 
                 {\cal O}_{n} \; d\sigma_n^{(2)} 
               + {\cal O}_{n} \circ d\alpha^{(0,2)}_{n}
               + {\cal O}_{n} \circ d\alpha^{(1,1)}_{n}.
 \nonumber
\end{eqnarray}
Here $d\alpha_{n+1}^{(0,1)}$ is a subtraction term for single unresolved configurations
of Born amplitudes.
This term is already known from NLO calculations.
The term $d\alpha_n^{(0,2)}$ is a subtraction term 
for double unresolved configurations.
Finally, $d\alpha_n^{(1,1)}$ is a subtraction term
for single unresolved configurations involving one-loop amplitudes.

To construct these terms the universal factorisation properties of 
QCD amplitudes in unresolved limits are essential.
QCD amplitudes factorise if they are decomposed into primitive
amplitudes.
\begin{figure*}
\begin{center}
\resizebox{160mm}{!}{
  \includegraphics[260pt,545pt][900pt,670pt]{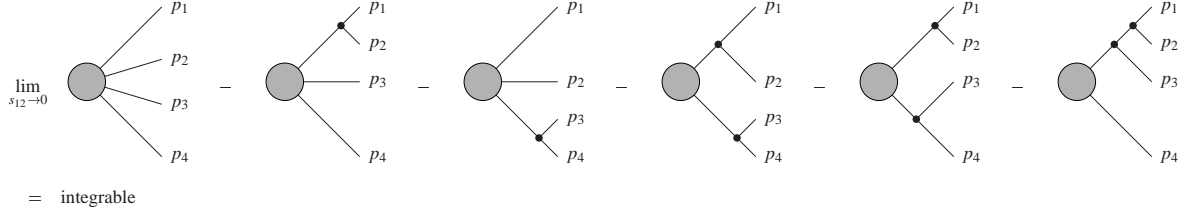}
}
\caption{Combinations of subtraction terms, which render the single unresolved limit
$s_{12} \rightarrow 0$ finite.}
\label{fig:1}
\end{center}
\end{figure*}
Primitive amplitudes are defined by
a fixed cyclic ordering of the QCD partons,
a definite routing of the external fermion lines through the diagram
and the particle content circulating in the loop.
One-loop amplitudes factorise in single unresolved limits as
\cite{Bern:1994zx,Bern:1998sc,Kosower:1999xi,Kosower:1999rx,Bern:1999ry,Catani:2000pi,Kosower:2003cz}
\begin{eqnarray}
\label{oneloopfactformula}
A^{(1)}_{n}
  & = &
  \mbox{Sing}^{(0,1)} 
  \cdot A^{(1)}_{n-1} +
  \mbox{Sing}^{(1,1)} \cdot A^{(0)}_{n-1}.
\end{eqnarray}
Tree amplitudes factorise in the double unresolved limits as
\cite{Berends:1989zn,Gehrmann-DeRidder:1998gf,Campbell:1998hg,Catani:1998nv,Catani:1999ss,DelDuca:1999ha,Kosower:2002su}
\begin{eqnarray}
\label{factsing}
A^{(0)}_{n}
  & = &
  \mbox{Sing}^{(0,2)} \cdot A^{(0)}_{n-2}.
\end{eqnarray}
To discuss the term $d\alpha_n^{(0,2)}$ let us consider as an example
the Born leading-colour contributions to $e^+ e^- \rightarrow q g g \bar{q}$,
which contribute to the NNLO corrections to
$e^+ e^- \rightarrow \mbox{2 jets}$.
The subtraction term has to match all double and single unresolved 
configurations.
It is convenient to construct $d\alpha_n^{(0,2)}$ as a sum over 
several pieces \cite{Weinzierl:2003fx},
\begin{eqnarray}
d \alpha^{(0,2)}_{n} & = & 
 \sum\limits_{\mbox{\tiny topologies $T$}} {\cal D}_{n}^{(0,2)}(T).
\end{eqnarray}
Each piece is labelled by a splitting topology.
Care has to be taken to disentangle correctly overlapping singularities,
such that the integrand is finite in all double and single unresolved limits.
Fig. \ref{fig:1} shows a combination which is finite in the single unresolved limit
$s_{12} \rightarrow 0$.
The integration over the double unresolved phase space involves square roots and leads to
new types of integrals with half-integer powers.
Mapping these integrals to sums, the new types are related to sums of the form
\cite{Weinzierl:2004bn}
\bq
\lefteqn{
\frac{1}{\Gamma\left(\frac{1}{2}\right)}
 \sum\limits_{n=1}^\infty 
 \frac{\Gamma\left(n+\frac{1}{2}\right)}{\Gamma(n+1)}
 \frac{x^n}{n^2}
 = 
 2 \left[ 
          \ln 2 \ln \left( 1+ \sqrt{1-x} \right)
 \right.
} & & \nonumber \\
& &  
 \left.
        + \ln 2 \ln \left( 1- \sqrt{1-x} \right)
        - \mbox{Li}_{11}\left(-\sqrt{1-x},1\right)
 \right. \nonumber \\
 & & \left.
        - \mbox{Li}_{11}\left(\sqrt{1-x},-1\right)
        - \mbox{Li}_{2}\left(-1\right)
        - \left( \ln 2 \right)^2
   \right],
 \nonumber \\
\lefteqn{
\Gamma\left(\frac{1}{2}\right)
 \sum\limits_{n=1}^\infty 
 \frac{\Gamma(n+1)}{\Gamma\left(n+\frac{1}{2}\right)}
 \frac{x^n}{n^2}
 = 
 - \mbox{Li}_{11}\left(\chi,1\right)
} & & \nonumber \\
 & &  
 + \mbox{Li}_{11}\left(\chi,-1\right)
 - \mbox{Li}_{11}\left(-\chi,1\right)
 + \mbox{Li}_{11}\left(-\chi,-1\right),
 \nonumber
\eq
where $\chi = \sqrt{-x/(1-x)}$.

\subsection{One-loop amplitudes with one unresolved parton}
\label{sect:oneloop}

Apart from $d\alpha_n^{(0,2)}$ also the term
$d\alpha_n^{(1,1)}$, which approximates one-loop amplitudes 
with one unresolved parton, is needed at NNLO.
If we recall the factorisation formula (\ref{oneloopfactformula}),
this requires as a new feature 
the approximation of the one-loop singular function
$\mbox{Sing}^{(1,1)}$.
The corresponding subtraction term is proportional to the 
one-loop $1\rightarrow 2$ splitting function 
${\cal P}^{(1,1)}_{(1,0)\; a \rightarrow b c}$.
An example is the leading-colour part for the
splitting $q \rightarrow q g$ \cite{Weinzierl:2003ra}:
\begin{eqnarray}
\lefteqn{
{\cal P}^{(1,1)}_{(1,0)\; q \rightarrow q g, lc, corr} 
 =  
     - \frac{11}{6\varepsilon} {\cal P}^{(0,1)}_{q \rightarrow q g}
 +
 S_\varepsilon^{-1} c_\Gamma \left( \frac{-s_{ijk}}{\mu^2} \right)^{-\varepsilon} 
 }
 \nonumber \\
 & &
  y^{-\varepsilon}
     \left\{ 
         g_{1, corr}(y,z) \; {\cal P}^{(0,1)}_{q \rightarrow q g}
         + f_2 \frac{2}{s_{ijk}} \frac{1}{y} p\!\!\!/_{e} 
 \right.
 \nonumber \\
 & & \left.
               \left[ 1 - \rho \varepsilon (1-y) (1-z) \right]
     \right\}
 \nonumber 
\end{eqnarray}
This term depends on the correlations among the remaining hard partons.
If only two hard partons are correlated, $g_{1}$ is given by
\bq
\lefteqn{
g_{1, intr}(y,z) 
 = 
  - \frac{1}{\eps^2} 
 \left[ \Gamma(1+\eps) \Gamma(1-\eps) \left( \frac{z}{1-z} \right)^\eps 
        + 1 
\right. } \nonumber \\ & & \left.
        - (1-y)^\eps z^\eps \; {}_2F_1\left( \eps, \eps, 1+\eps; (1-y)(1-z) \right) \right].
\hspace{10mm} 
\nonumber
\eq
Here, $y=s_{ij}/s_{ijk}$, $z=s_{ik}/(s_{ik}+s_{jk})$ and 
$f_2=(1-\rho\varepsilon)/2/(1-\varepsilon)/(1-2\varepsilon)$.
The parameter $\rho$ specifies the variant of dimensional regularisation:
$\rho  = 1$ in the conventional or 't Hooft-Veltman scheme and 
$\rho=0$ in a four-dimensional scheme.
For the integration of the subtraction terms 
over the unresolved phase space all occuring integrals are reduced to
standard integrals of the form
\bq
\lefteqn{
\int\limits_0^1 dy \; y^a (1-y)^{1+c+d} \int\limits_0^1 dz \; z^c (1-z)^d \left[ 1 -z(1-y)\right]^e
} \nonumber \\
\lefteqn{
  {}_2F_1\left( \eps, \eps; 1+\eps; (1-y) z \right)
 = }
 \nonumber \\ & &
 \frac{\Gamma(1+a) \Gamma(1+d) \Gamma(2+a+d+e) \Gamma(1+\eps)}{\Gamma(2+a+d) \Gamma(\eps) \Gamma(\eps)}
 \nonumber \\ & &
 \sum\limits_{j=0}^\infty 
 \frac{\Gamma(j+\eps) \Gamma(j+\eps) \Gamma(j+1+c)}
      {\Gamma(j+1) \Gamma(j+1+\eps) \Gamma(j+3+a+c+d+e)}.
 \nonumber
\eq
The result is proportional to  hyper-geometric functions 
${}_4F_3$ with unit argument and can be
expanded into a Laurent series in $\varepsilon$ 
with the techniques of \cite{Moch:2001zr,Weinzierl:2002hv}.
For the example discussed above one finds after integration
\cite{Weinzierl:2003ra}:
\bq
\lefteqn{
{\cal V}^{(1,1)}_{(1,0)\; q \rightarrow q g, lc, intr} = 
 - \frac{1}{4\eps^4}
 - \frac{31}{12 \eps^3}
 + \left( -\frac{51}{8} - \frac{1}{4} \rho 
\right. }
 \nonumber \\
 & & \left.
 + \frac{5}{12} \pi^2 
          - \frac{11}{6} L
   \right) \frac{1}{\eps^2} 
 + \left( - \frac{151}{6} - \frac{55}{24} \rho 
          + \frac{145}{72} \pi^2 
 \right. \nonumber \\ & & \left.
          + \frac{15}{2} \zeta_3
          - \frac{11}{4} L 
          - \frac{11}{12} L^2 
   \right) \frac{1}{\eps}
 - \frac{1663}{16} - \frac{233}{24} \rho 
 \nonumber \\ & & 
 + \frac{107}{16} \pi^2 + \frac{5}{12} \rho \pi^2 
 + \frac{356}{9} \zeta_3 
 - \frac{1}{72} \pi^4
 - \frac{187}{24} L 
 \nonumber \\
 & &
 - \frac{11}{12} \rho L 
 + \frac{55}{72} \pi^2 L
 - \frac{11}{8} L^2 - \frac{11}{36} L^3
 + i \pi \left[
            - \frac{1}{4 \eps^3}
 \right. \nonumber \\ & & \left.
            - \frac{3}{4 \eps^2}
            + \left( - \frac{29}{8}
                     - \frac{1}{4} \rho + \frac{\pi^2}{3} \right) \frac{1}{\eps}
            - \frac{139}{8} - \frac{11}{8} \rho 
 \right. \nonumber \\
 & & \left.
            + \pi^2 + \frac{15}{2} \zeta_3 
     \right]
 + {\cal O}(\eps),
 \nonumber
\eq
where $L = \ln(s_{ijk}/\mu^2)$.

\section{Outlook}
\label{sec:outlook}

In this talk I discussed methods for the cancellation of infrared singularities at NNLO.
The handling of these divergences is the remaining bottleneck in the construction of fully
differential numerical programs at NNLO.
With the progress we witnessed in the field in the last years we can expect to obtain numerical results rather soon.
An example would be the extension of 
exisiting numerical programs for NLO predictions
on $e^+ e^- \rightarrow 4 \;\mbox{jets}$ 
\cite{Dixon:1997th,Nagy:1998bb,Campbell:1998nn,Weinzierl:1999yf}
towards NNLO predictions
for $e^+ e^- \rightarrow 3 \;\mbox{jets}$.

{\bf Acknowledgments:} I would like to thank the organizers for a
stimulating conference Loops and Legs 2004.


\end{document}